\documentclass{article}
\usepackage{spconf,amsmath,graphicx,hyperref}
\usepackage{booktabs}
\usepackage{booktabs} 
\usepackage{tabularx} 
\usepackage{multirow} 
\usepackage{capt-of} 
\usepackage{float} 
\usepackage{amssymb}

\title{WEE-Therapy: A Mixture of Weak Encoders Framework for
Psychological Counseling Dialogue Analysis}
\twoauthors
  {Yongqi Kang\thanks{}}
  {Sichuan Unversity\\
  Department of Computer Science}
  {Yong Zhao\thanks{}}
  {Sichuan Unversity\\
  Department of Computer Science
  }
\begin{document}
%
\maketitle
\begin{abstract}
The advancement of computational psychology requires AI tools capable of deeply understanding counseling dialogues. Existing audio language models (AudioLLMs) often rely on single speech encoders pre-trained on general data, struggling to capture domain-specific features like complex emotions and professional techniques. To address this, we propose \textbf{WEE-Therapy}, a multi-task AudioLLM incorporating a \textbf{Weak Encoder Ensemble (WEE)} mechanism. This supplements a powerful base encoder with a pool of lightweight, specialized encoders. A novel dual-routing strategy combines stable, data-independent domain knowledge with dynamic, data-dependent expert selection. Evaluated on emotion recognition, technique classification, risk detection, and summarization, WEE-Therapy achieves significant performance gains across all tasks with minimal parameter overhead, demonstrating strong potential for AI-assisted clinical analysis.
\end{abstract}

\section{Introduction}
Mental health is a core pillar of human well-being, and psychological counseling, as a crucial safeguard, faces multiple challenges such as resource shortages, high supervision costs, and subjective analysis methods. In recent years, breakthrough advancements in artificial intelligence, particularly in natural language processing (NLP)—especially the powerful dialogue and reasoning capabilities demonstrated by large language models (LLMs)~\cite{devlin2019bert}—have provided a new paradigm for developing computational tools. Among these, audio language models (AudioLLMs), which can directly understand raw speech rich in paralinguistic information (such as tone, pauses, and emotions), are particularly well-suited for in-depth analysis of psychological counseling dialogues. They hold the potential to enable objective quantification of dialogue processes, automatic identification of intervention techniques, and timely warnings of high-risk moments.

However, directly applying advanced AudioLLMs to the \textbf{highly specialized domain} of psychological counseling reveals a significant ``domain adaptation'' gap. Existing mainstream methods typically rely on large speech encoders (e.g., Whisper~\cite{radford2023robust}) pre-trained on general corpora (e.g., LibriSpeech), whose representational spaces are not optimized for capturing \textbf{domain-specific features} in psychological counseling. Counseling dialogues are filled with complex emotional fluctuations, specific professional terminology, subtle turn-taking, and silences and sighs that carry critical information—nuances that are difficult for general-purpose encoders to fully capture. Although scaling up the model or conducting comprehensive domain-specific pre-training could mitigate this issue, these approaches face significant obstacles in terms of data acquisition, computational costs, and deployment feasibility.

A promising solution is the Mixture of Weak Encoders (MoWE) architecture~\cite{zhang2025mowe}. Instead of seeking a single ``all-powerful'' giant encoder, this approach employs a powerful base encoder supplemented by a set of lightweight ``expert'' encoders, dynamically integrating their features through a routing mechanism. This architecture has already demonstrated its ability to efficiently expand model capabilities in general audio tasks. However, \textbf{its effectiveness, adaptation methods, and potential value in specialized domains such as psychological counseling remain an unexplored open question}.

To address this research gap, this paper introduces \textbf{WEE-Therapy}, a multi-task AudioLLM framework specifically tailored for psychological counseling analysis. Our core idea is to \textbf{leverage domain knowledge-driven integration and adaptation to transform the existing WEE paradigm into an effective tool for addressing domain-specific challenges}. Specifically, the main contributions of this study are as follows:

\begin{itemize}
    \item \textbf{Pioneering Domain Application}: To the best of our knowledge, this is the \textbf{first} systematic application of the MoWE architecture in the field of computational psychology, providing a novel and efficient solution to the domain adaptation challenges in psychological counseling analysis.
    \item \textbf{Domain-Adapted Design}: Rather than simply reusing existing models, we made key adaptations based on domain insights. These include constructing a mixed pool integrating an \textbf{emotion expert encoder} and designing a dual-routing strategy that incorporates \textbf{domain priors} to ensure stable extraction of psychologically critical features.
    \item \textbf{Systematic Empirical Evaluation}: We built a comprehensive analysis system and conducted a thorough evaluation across four core tasks: emotion recognition, counseling technique classification, crisis risk detection, and dialogue summarization. The experimental results not only validate the effectiveness of the framework but also provide in-depth insights into the specialized behaviors of different ``expert'' encoders, offering valuable references for future research.
    \item \textbf{Application Value Orientation}: This study clearly highlights the \textbf{significant application potential} of the proposed framework for developing low-cost, high-efficiency AI-assisted clinical supervision and analysis systems. The methodology also offers broad implications for adapting general large models to other vertical domains.
\end{itemize}

The remainder of this paper is structured as follows: Section 2 reviews the WEE-THERAPY framework; Section 3 elaborates on the experimental setup; Section 4 got conclusion;

\section{The WEE-Therapy Framework}
\label{sec:format}
This section will elaborate in detail on the proposed \textbf{WEE-Therapy} framework. The core idea of this framework is to enhance the base audio language model (AudioLLM) through a \textbf{WEE} module, enabling it to better adapt to the complexity and specificity of psychological counseling dialogues. The overall architecture of the system is illustrated in Figure 1.
\begin{figure}[htbp]
    \centering
    \includegraphics[width=0.95\linewidth]{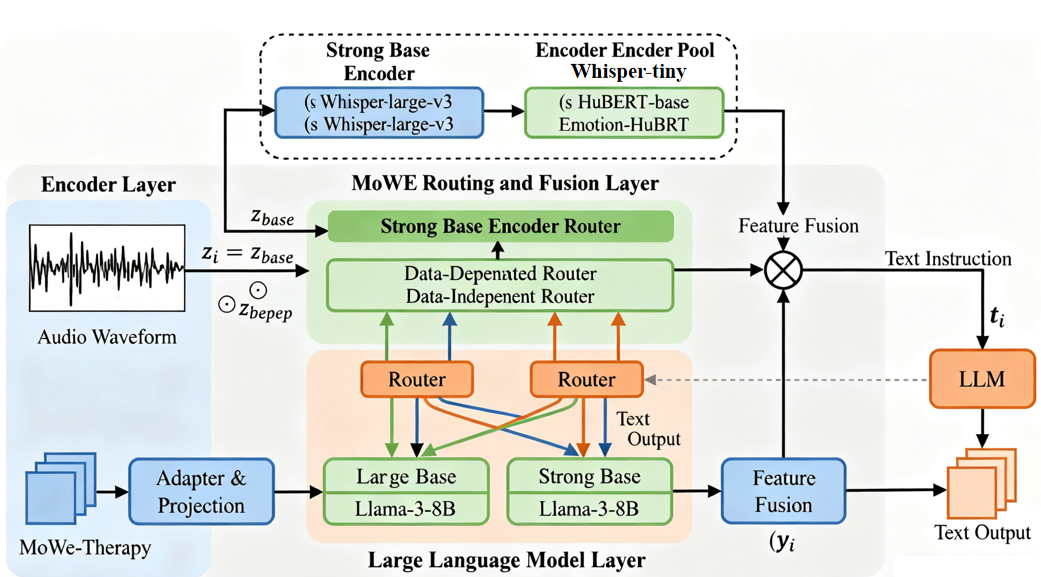} 
    \caption{Overall architecture of the proposed MoWe-Therapy framework.}
    \label{fig:arch}
\end{figure}
\subsection{Overall Framework (WEE-Therapy Framework)}

Our framework primarily consists of the following three core components:

\begin{enumerate}
    \item \textbf{Encoder Layer}: Responsible for converting the input raw counseling dialogue audio into high-dimensional feature representations.
    \begin{itemize}
        \item \textbf{Strong Base Encoder (\(E_{base}\))}: We employ a large-scale, high-performance general-purpose speech encoder as the backbone, such as \textbf{Whisper-large-v3}~\cite{radford2023robust}. This encoder has a substantial number of parameters ($\sim$637M) and excels in general speech recognition tasks, providing us with a stable and powerful base audio representation \(z_{base} = E_{base}(a_i)\), where \(a_i\) represents the \(i\)-th input audio segment.
        \item \textbf{Weak Encoder Pool (\(\{E_k\}_{k=1}^M\))}: To supplement the fine-grained features that the base encoder might miss in the vertical domain of psychological counseling, we introduce a pool of \(M\) lightweight encoders. These ``weak'' encoders have significantly fewer parameters than the base encoder (typically an order of magnitude less), such as \textbf{HuBERT-base}~\cite{hsu2021hubert}, \textbf{Wav2Vec2.0-base}~\cite{baevski2020wav2vec}, or specialized encoders fine-tuned on emotion datasets (e.g., IEMOCAP~\cite{busso2008iemocap}). They each have their own strengths, collectively forming a flexible ``committee of experts.''
    \end{itemize}

    \item \textbf{WEE Routing and Fusion Layer}: This is the innovative core of this work. This layer contains a \textbf{Router}, whose function is to intelligently select and activate the most relevant subset from the weak encoder pool based on the input audio. Specifically, we designed a \textbf{dual-routing strategy} (detailed in Section 3.2), which generates both data-dependent and data-independent weak encoder features \(z_{dep}\) and \(z_{indep}\). Subsequently, these weak encoder features are concatenated (Concatenation) with the base encoder features along the feature dimension to form the final enhanced audio representation:
    \[
    z_i = z_{i,base} \oplus_f z_{i,MoWE} = z_{i,base} \oplus_f (z_{i,dep} \oplus_f z_{i,indep})
    \]
    This approach greatly enriches the information content of the input features without increasing the sequence length (and thus without significantly increasing the computational burden on the LLM).

    \item \textbf{Large Language Model Layer}:
    \begin{itemize}
        \item \textbf{Adapter \& Projection}: Since the output embedding dimensions of the audio encoder typically do not match the input space of the LLM, we use a lightweight adapter (e.g., a linear layer plus GELU activation) to perform downsampling on the concatenated features \(z_i\), and then map them to the LLM's token embedding space via a projection layer, generating audio tokens \(token_{a_i} = proj(adapter(z_i))\).
        \item \textbf{Text Generation}: Simultaneously, the text instruction \(t_i\) (e.g., ``Analyze the counselor's techniques in this dialogue'') is converted into text tokens \(token_{t_i}\) via the LLM's tokenizer. The audio tokens and text tokens are concatenated along the sequence dimension and fed into a large language model (e.g., \textbf{Llama-3-8B-Instruct}~\cite{touvron2023llama}). The LLM generates the analysis result in text form \(\hat{y}_i = LLM([token_{a_i}; token_{t_i}])\) in an autoregressive manner using prefix-conditioned generation.
    \end{itemize}
\end{enumerate}
\subsection{Weak Encoder Ensemble}
The workflow of the dual-routing strategy in the MoWE module is as follows:

\begin{itemize}
    \item \textbf{Data-Independent Router}: The goal of this router is to select a \textbf{fixed} weak encoder that provides a \textbf{global, content-agnostic} supplement of domain knowledge for \textbf{every} input sample. For example, it might always prefer the encoder fine-tuned on emotional data to ensure that the emotional features of all counseling dialogues are enhanced. Its computation process is as follows:
    \[
    r_{indep} = \text{KeepTop1}(\text{Softmax}(w_{indep}))
    \]
    \[
    z_{i,indep} = \sum_{k=1}^{M} r_{indep}[k] \cdot E_k(a_i)
    \]
    Here, \(w_{indep} \in \mathbb{R}^M\) is a learnable parameter vector, which can be initialized with priors (e.g., setting a higher initial value for the emotion encoder). \(\text{KeepTop1}\) is an operator that returns a one-hot vector where only the position with the highest weight is 1, and the others are 0.

    \item \textbf{Data-Dependent Router}: The goal of this router is to act as an ``on-site conductor,'' dynamically selecting the most appropriate ``on-site expert'' \textbf{based on the specific content of the current input audio}. Its decision relies on the global audio features extracted by the base encoder. The computation process is as follows:
    \[
    \bar{z}_{i,base} = \text{MeanPool}(z_{i,base}) \quad 
    \]
    \[
    r_{i,dep} = \text{KeepTop1}(\text{Softmax}(\bar{z}_{i,base} W_{dep}))
    \]
    \[
    z_{i,dep} = \sum_{k=1}^{M} r_{i,dep}[k] \cdot E_k(a_i)
    \]
    Here, \(W_{dep} \in \mathbb{R}^{d_{base} \times M}\) is a learnable projection matrix that maps the features of the base encoder to a routing score space corresponding to the number of weak encoders \(M\).
\end{itemize}

\subsection{Training Objective}
The model is trained using a multi-task learning paradigm. The total loss function consists of two parts:

\[
\mathcal{L} = \mathcal{L}_{\text{next-token}} + \lambda \cdot \mathcal{L}_{\text{MoWE}} \quad (\lambda=0.1)
\]

\begin{enumerate}
    \item \textbf{Next-Token Prediction Loss (\(\mathcal{L}_{\text{next-token}}\))}: This is the standard autoregressive loss for training the LLM, i.e., maximizing the likelihood of the target response sequence.

    \item \textbf{WEE Routing Loss (\(\mathcal{L}_{\text{WEE}}\))}: To train the routers to make good and balanced decisions, we design a specialized auxiliary loss:
    \[
    \mathcal{L}_{\text{WEE}} = \frac{1}{2} [\mathcal{L}_{\text{indep-ent}} + (\mathcal{L}_{\text{dep-ent}} + \mathcal{L}_{\text{dep-div}})]
    \]
    \begin{itemize}
        \item \textbf{Entropy Loss (\(\mathcal{L}_{\text{indep-ent}}\) \& \(\mathcal{L}_{\text{dep-ent}}\))}: Encourages the router to make ``confident'' decisions, i.e., producing a sharper output distribution.
        \[
        \mathcal{L}_{\text{indep-ent}} = -\sum_{k=1}^{M} r_{\text{indep}}[k] \cdot \log(r_{\text{indep}}[k])
        \]
        \[
        \mathcal{L}_{\text{dep-ent}} = -\frac{1}{B} \sum_{i=1}^{B} \sum_{k=1}^{M} r_{i,\text{dep}}[k] \cdot \log(r_{i,\text{dep}}[k])
        \]
        \item \textbf{Diversity Loss (\(\mathcal{L}_{\text{dep-div}}\))}: Prevents the data-dependent router from always selecting the same encoder, encouraging the utilization of all weak encoders.
        \[
        \bar{r}_{\text{dep}} = \frac{1}{B} \sum_{i=1}^{B} r_{i,\text{dep}}
        \]
        \[
        \mathcal{L}_{\text{dep-div}} = \sum_{k=1}^{M} \bar{r}_{\text{dep}}[k] \cdot \log(\bar{r}_{\text{dep}}[k])
        \]
    \end{itemize}
    Here, \(B\) is the training batch size.
\end{enumerate}

During training, we freeze most parameters of the base encoder and the LLM. We primarily fine-tune the routing networks, the adapter, the projection layer, and a small number of trainable parameters injected into the LLM via LoRA (Low-Rank Adaptation)~\cite{hu2022lora}. This is an efficient parameter fine-tuning strategy that effectively prevents overfitting.

\section{Experimental Setup}
To comprehensively evaluate the effectiveness of our proposed \textbf{WEE-Therapy} framework, we designed multi-task experiments and conducted tests on several representative psychological counseling datasets. This section elaborates in detail on the tasks and datasets used in the experiments, evaluation metrics, model implementation details, and training configurations.
\label{sec:pagestyle}

\subsection{Tasks and Datasets}
We selected four tasks that comprehensively reflect the core requirements of psychological counseling analysis. Due to the sensitivity of psychological counseling data, publicly available datasets are limited. Our experiments are partially based on existing public datasets and partially on simulated data. Table~\ref{tab:datasets} summarizes the tasks, datasets, and evaluation metrics used in the experiments.

\begin{table}[H] 
\centering
\small
\setlength{\tabcolsep}{4pt}
\caption{Summary of Experimental Tasks, Datasets, and Evaluation Metrics}
\label{tab:datasets}
\begin{tabular}{p{0.18\columnwidth}p{0.22\columnwidth}p{0.55\columnwidth}}
\toprule
\textbf{Task} & \textbf{Dataset} & \textbf{Description \& Metric} \\
\midrule
\textbf{Emotion Recognition (ER)} & \textbf{DAIC-WOZ}~\cite{burdisso2024daic} & 
\begin{minipage}[t]{\linewidth}
  \textbf{Description}: Audio recordings of clinical diagnostic interviews, annotated with psychological distress states (e.g., anxiety, depression). Used for emotion state classification.\\
  \textbf{Metric}: \textbf{Macro F1-Score}, with focus on negative emotion recognition (anxiety, depression).
\end{minipage} \\
\midrule
\textbf{Counselor Technique Classification (CTC)} & \textbf{Simulated Dataset} & 
\begin{minipage}[t]{\linewidth}
  \textbf{Description}: Simulated dataset with counseling technique labels (Questioning, Empathizing, Restating, Affirming, etc.).\\
  \textbf{Metric}: \textbf{Accuracy} and \textbf{Macro F1-Score}.
\end{minipage} \\
\midrule
\textbf{Crisis Risk Detection (CMD)} & \textbf{Self-Annotated} & 
\begin{minipage}[t]{\linewidth}
  \textbf{Description}: Identifies high-risk moments revealing suicidal or self-harm intentions. Fine-grained annotations on data segments.\\
  \textbf{Metric}: \textbf{Precision@K} (due to extreme class imbalance).
\end{minipage} \\
\midrule
\textbf{Dialogue Summarization (DS)} & \textbf{Self-Annotated} & 
\begin{minipage}[t]{\linewidth}
  \textbf{Description}: Generates concise summaries capturing core content, client issues, and intervention strategies.\\
  \textbf{Metric}: \textbf{ROUGE-L}~\cite{lin2004rouge} score.
\end{minipage} \\
\midrule
\textbf{Overall Judgment} & \textbf{All Datasets} & 
\begin{minipage}[t]{\linewidth}
  \textbf{Metric}: \textbf{GPT-4 as judge} providing 0-5 score based on alignment, professionalism, and completeness.
\end{minipage} \\
\bottomrule
\end{tabular}
\end{table}
\vspace{-\baselineskip} 
\begin{table}[t]
\centering
\footnotesize
\caption{Main results on counseling analysis tasks. Best results are \textbf{bold}. $\Delta$ shows improvement over Base.}
\label{tab:main_results}
\begin{tabular}{l c c c c}
\toprule
\textbf{Model} & \textbf{ER (F1)} & \textbf{CTC (Acc)} & \textbf{CMD (P@5)} & \textbf{DS (R-L)} \\
\midrule
Whisper-only & 67.2 & 73.5 & 72.1 & 31.6 \\
HuBERT-only & 65.9 & 70.1 & 68.5 & 29.8 \\
\midrule
Data-Indep. only & 69.0 & 75.2 & 75.3 & 33.1 \\
Data-Dep. only & 70.5 & 76.9 & 77.8 & 34.9 \\
\textbf{WEE(Ours)} & \textbf{72.6} & \textbf{78.9} & \textbf{80.1} & \textbf{36.8} \\
\midrule
$\Delta$ & \textbf{+5.4} & \textbf{+5.4} & \textbf{+8.0} & \textbf{+5.2} \\
\bottomrule
\end{tabular}
\end{table}
All audio inputs were uniformly cropped or padded to \textbf{30 seconds} during preprocessing and resampled to 16kHz to meet the input requirements of the encoders.

\subsection{Implementation Details}

\begin{itemize}
    \item \textbf{Base Models}:
    \begin{itemize}
        \item \textbf{Strong Base Encoder}: We adopted \textbf{Whisper-large-v3}~\cite{radford2023robust} ($\sim$637M parameters) as the default strong base encoder. Its strong performance on general speech tasks provides a solid foundation for our system.
        \item \textbf{Large Language Model}: We primarily used \textbf{Llama-3-8B-Instruct}~\cite{touvron2023llama} as the core backbone for text generation. To validate the generality of the method, we also conducted supplementary experiments on \textbf{Zephyr-7B}~\cite{tunstall2023zephyr} and \textbf{Phi-3-mini-4k-instruct}~\cite{abdin2024phi} (3.8B parameters).
    \end{itemize}

    \item \textbf{Weak Encoder Pool}:
    \begin{itemize}
        \item In our main experiments, the WEE pool contained 3 weak encoders to balance performance and efficiency:
        \begin{enumerate}
            \item \textbf{Whisper-tiny}~\cite{radford2023robust} (39M parameters): A lightweight general-purpose speech encoder that provides efficient speech content perception.
            \item \textbf{HuBERT-base}~\cite{hsu2021hubert} (95M parameters): Trained based on self-supervised learning, it excels at learning discrete representations of speech and is sensitive to phonemes and acoustic content.
            \item \textbf{Emotion-Finetuned-HuBERT} (95M parameters): An encoder obtained by fine-tuning HuBERT-base on the \textbf{IEMOCAP}~\cite{busso2008iemocap} emotion recognition dataset, specifically designed to capture emotional features.
        \end{enumerate}
        
    \end{itemize}


\end{itemize}

\section{Conclusion}
\label{sec:Con}

We proposed \textbf{WEE-Therapy}, a parameter-efficient framework for psychological counseling analysis. To address \textbf{task diversity} and \textbf{data scarcity}, our model employs a mixture-of-experts architecture with a novel WEE Routing mechanism that dynamically combines strong and weak encoders.

Trained with a specialized routing loss, our framework efficiently learns to perform multiple counseling tasks—including emotion recognition, technique classification, risk detection, and dialogue summarization—while minimizing overfitting through selective parameter tuning.

Experimental results demonstrate that WEE-Therapy effectively handles the nuanced requirements of counseling dialogue analysis. Our work provides a promising foundation for developing AI assistants that can enhance mental health support services.

Future work will explore the framework's application to larger-scale real-world counseling datasets and other low-resource domains.

\bibliographystyle{IEEEbib}
\bibliography{strings,refs}

\end{document}